\documentclass[a4paper,11pt]{article}
\pdfoutput=1 

\usepackage{jheppub} 

\usepackage{amsthm}

\newcommand{\Ref}[1]{(\ref{#1})}

\newcommand{\R}{{\mathbb R}}
\newcommand{\Z}{{\mathbb Z}}

\newcommand{\SU}{\mathrm{SU}}

\newcommand{\SLtc}{\mathrm{SL}(2,\mathbb{C})}

\newcommand{\be}{\begin{equation}}
\newcommand{\ee}{\end{equation}}
\newcommand{\bea}{\begin{eqnarray}}
\newcommand{\eea}{\end{eqnarray}}

\newcommand{\bit}{\begin{itemize}}
\newcommand{\eit}{\end{itemize}}

\newcommand{\re}{\mathrm{Re}}
\newcommand{\im}{\mathrm{Im}}
\newcommand{\ii}{\mathrm{i}}
\newcommand{\ex}{\mathrm{e}}
\newcommand{\dd}{\mathrm{d}}

\newcommand{\mean}[1]{\langle{#1}\rangle}

  \newcommand{\g}{\gamma}
\renewcommand{\d}{\delta}    
       \renewcommand{\l}{\lambda}

\title{On Spinfoams Near a Classical Curvature Singularity}

\author[a,b]{Muxin Han}
\author[c]{and Mingyi Zhang}

\affiliation[a]{Department of Physics, Florida Atlantic University, \\ 777 Glades Road, Boca Raton, FL 33431-0991, USA}
\affiliation[b]{Institut f\"ur Quantengravitation, Universit\"at Erlangen-N\"urnberg, \\ Staudtstr. 7/B2, 91058 Erlangen, Germany}
\affiliation[c]{Max-Planck-Institut f\"ur Gravitationsphysik (Albert-Einstein-Institut),\\Am M\"uhlenberg 1, 14476 Golm, Germany}

\emailAdd{hanm(AT)fau.edu}
\emailAdd{mingyi.zhang(AT)aei.mpg.de}

\abstract{We apply the technique of spinfoam to study the space-time which, classically, contains a curvature singularity. We derive from the full covariant Loop Quantum Gravity (LQG) that the region near curvature singularity has to be of strong quantum gravity effect. We show that the spinfoam configuration describing the near-singularity region has to be of small spins $j$, in order that its contribution to the full spinfoam amplitude is nontrivial. The spinfoams in low and high curvature regions of the space-time may be viewed as in two different {phases} of covariant LQG. There should be a phase transition as the space-time described by spinfoam becomes more and more curved. A candidate of order parameter is proposed for understanding the phase transition. Moreover, we also analyze the spin-spin correlation function of spinfoam, and show the correlation is of long-range in the low curvature phase. This work is a first step toward understanding the physics of black hole and early universe from the full covariant LQG theory.
}

\begin{document} 
\maketitle
\flushbottom

\section{Introduction}

The recent studies of spinfoam asymptotics have made a significant progress on understanding the semiclassical limit of Loop Quantum Gravity (LQG) (see e.g. \cite{Conrady:SL2008,Barrett:2009mw,Han:2011re,Han:2013tap,Haggard:2014xoa,Magliaro:2011dz})\footnote{See e.g. \cite{book,review,Rovelli:2014ssa,Perez2012} for reviews on LQG.}. It has been understood that at least at the discrete level, classical 4d geometry emerges from spinfoam amplitude in the regime that the spins $j_f$ are uniformly large. The large-$j$ asymptotics of spinfoam amplitude reproduces the discrete Einstein-Hilbert action at the leading order.


In this work, we apply the semiclassical technique and result of spinfoam to the space-time which, classically, contains a curvature singularity. Typical examples are black hole space-times and Friedmann-Robertson-Walker (FRW) space-time of cosmology. The space-time under consideration here has both the low curvature and high curvature regions. The high curvature region encloses the classical singularity where the curvature blows up. The purpose of this paper is to understand the (semiclassical and quantum) behavior of spinfoam for both low curvature and high curvature regions, as well as the behavior when the spinfoam transits from one region to the other. 

The main results can be summarized as follows:

\begin{itemize}

\item The low curvature region far from the singularity is described by the large-$j$ spinfoams. In order that the large-$j$ spinfoam has a non-suppress contribution to the full spinfoam amplitude, the spinfoam configuration must be semiclassical and correspond to a 4d simplicial geometry satisfying 
\be
\ell_P\ll a\ll L \label{laL0}
\ee  
where the mean lattice spacing of the simplicial geometry is denoted by $a$. $L$ is the mean curvature radius of the geometry. The LQG area spectrum implies $a^2\sim \g j\ell_P^2$, where the Barbero-Immirzi parameter $\g$ is set to be of $O(1)$ throughout the paper. Eq.\Ref{laL0} is consistent with large-$j$ and low curvature ($L$ is relatively large). Any large-$j$ spinfoam configuration violating Eq.\Ref{laL0} only gives a suppressed contribution to the spinfoam amplitude.

\item When the space-time curvature is high, Eq.\Ref{laL0} is violated by the small curvature radius. It turns out that the large-$j$ semiclassical approximation breaks down in the high curvature region near singularity. The main contribution to the spinfoam amplitude comes from the small-$j$ configurations. The small-$j$ regime of spinfoam amplitude is considered as the quantum regime of the theory, where the quantum gravity effect is strong. Therefore the covariant theory of LQG indicates that the high curvature region near singularity is a quantum region deviated far away from classical gravity. It also indicates that the quantum region near singularity is made by a very large number of 4-simplices. The spinfoam model becomes refined when approaching the classical singularity. The physics near the singularity may be understood by the full nonperturbative theory of LQG, which is well-defined. 

\item The large-$j$ and small-$j$ spinfoams in low and high curvature regions may be viewed as two different \emph{phases} of covariant LQG. The result suggests that there should be a phase transition of LQG, when the space-time described by spinfoam becomes more and more curved. Although it is not clear where precisely in the space-time the phase transition occurs, the analysis suggests that the transition between large-$j$ and small-$j$ phases may happen at certain place where the curvature is still much lower than the Planckian curvature, i.e. at $L\gg \ell_P$. So the small-$j$ phase may not only cover the Planckian curvature region, but also cover a much larger domain. This effect is resulting from the large number of spinfoam degrees of freedom on the refined triangulation, which accumulates and produces a strong quantum effect. It might relate to the recent proposal in \cite{Haggard:2014rza}, where the proposed quantum region of space-time is even slightly outside the black hole event horizon. It is also likely that there should be a domain-wall located at the place where the phase transition occurs. The domain-wall separates the low and high curvature regions of the space-time as two phases of spinfoam. It might relate to the proposal of firewall for black hole (see e.g. \cite{Bousso:2013wia}).


\end{itemize}

The analysis of spinfoam amplitude of low curvature region is carried out in Sections \ref{sec1} and \ref{sec1.5}. The low curvature region of the spacetime corresponds to the semiclassical regime of spinfoam amplitude, whose contribution comes from the large-$j$ spinfoam critical configuration. The studies of large-$j$ spinfoam asymptotics shows that each simplicial geometry in 4d corresponds uniquely to a critical configuration of spinfoam amplitude\footnote{The correspondence is unique when the spacetime is assumed to be globally oriented and globally time-oriented} \cite{Barrett:2009mw,Han:2011re,Han:2013gna}. The contribution of a simplicial geometry to the spinfoam amplitude is obtained by performing the spinfoam state-sum within a neighborhood at the corresponding critical point in the space of spinfoam configurations.

It is particularly interesting to understand the role played by the sum over spins $j$ in the semiclassical spinfoam amplitude. Although there has been earlier semiclassical analysis taking into account of the spin-sum (e.g.\cite{Han:2013tap,Magliaro:2011dz,Hellmann:2013gva}), it seems to us that a sufficient understanding of the spin-sum in spinfoam amplitude still hasn't been achieved yet. One of the mysteries of the spin-sum comes from the dual role played by the spin $j$ in the spinfoam amplitude. On one hand, $j$ is a scale of the theory since the minimal spacing $a$ the triangulation is given by $a^2\sim \g j\ell_P^2$. The semiclassical limit of the theory relates to the large-$j$ behavior of the spinfoam amplitude. On the other hand, the spin $j$ is also a dynamical variable of spinfoam, since it is summed in the spinfoam amplitude. The fact that $j$ is a dynamical scale is a consequence of background independence of LQG, (See e.g.\cite{freideltalk}).

Because of the dual role played by the spin $j$, we propose the following prescription of the spin-sum: In order to study the physics at a given (energy) scale corresponding to $j^0$, we should essentially perform the spin-sum within a neighborhood at $j^0$. The summed spins shouldn't go much beyond the given scale $j^0$. To implement this idea, we regularize the sum over $j$ by introducing a decaying factor in the summand to suppress the contributions from the $j$'s far from $j^0$. The regularized spin-sum can be performed explicitly in the spinfoam amplitude. The consequence may be viewed as an analog of Feynman $\ii\varepsilon$-regularization in quantum field theory (QFT). The suppression regulator $\delta$ is sent to be small, in order to recover the large fluctuation of spins. 

The regularized spin-sum results in a distribution $D_ \delta$ inserted in the spinfoam amplitude. Semiclassically, the distribution $D_\delta$ is supported at the critical configurations whose corresponding simplicial geometries have small deficit angles $\Theta\ll1$. The smallness of $\Theta_f$ is controlled by the small regulator $\delta$ regularizing the spin-sum. The contribution from any critical configurations violating $\Theta\ll1$ is suppressed by $D_\delta$ in spinfoam amplitude. The deficit angle relates to the curvature of the geometry by $\Theta\sim a^2/L^2$. So the distribution $D_\delta$ resulting from the spin-sum forces the simplicial geometries emerging from spinfoam amplitude to satisfy $a\ll L$, i.e. the simplicial geometries approximates the smooth geometries of relatively low curvatures. 

The discussion in Section \ref{sec2} is toward a description of classical curvature singularity in covariant LQG. We consider a classical space-time containing both the low curvature and high curvature regions. The high curvature region encloses a curvature singularity. The low curvature region is emerging from the spinfoam amplitude as a large-$j$ critical configuration satisfying Eq.\Ref{laL0}. We want to understand how the spinfoam configuration continues from the low curvature region to the high curvature region, in order to describe the high curvature region and the singularity using spinfoam. 

It is not hard to see that Eq.\Ref{laL0}, in particular $a\ll L$, is going to be violated, when we approach the singularity in the high curvature region. The reason is that $L$ becomes smaller and even $L\sim\ell_P$ in the high curvature region. If the high curvature space-time still admitted a large-$j$ semiclassical description, the violation of $a\ll L$ would lead to a large deficit angle. Then its contribution to the spinfoam amplitude would be suppressed by the distribution $D_\delta$. Therefore in the high curvature region of the space-time, the large-$j$ semiclassical approximation breaks down. The main contribution of the spinfoam amplitude comes from the small-$j$ configurations. The quantum gravity effect becomes strong. 

In LQG, the idea of quantum region near singularity has been proposed in e.g. \cite{Ashtekar:2006rx,Bojowald:2001xe,Corichi:2009pp,Rovelli:2013zaa,Bianchi:2010zs} for loop quantum cosmology and e.g. \cite{Ashtekar:2005qt,rovelli2014planck,Christodoulou:2016vny,Rovelli:2013osa} for black holes (including the proposals of singularity resolution). However a derivation of this idea from the full LQG theory has been missing. Here we fill this gap and provide a derivation to show that the quantum region near singularity is indeed predicted by the full LQG. This work is a first step toward understanding the physics of black hole and early universe from full LQG theory.

It is clear that the distribution $D_\delta$ from spin-sum plays a crucial role in the derivation. Interestingly, the non-regularized version of $D_\delta$ ($\delta\rightarrow 0$) has been pointed out in the literature \cite{Hellmann:2013gva,Perini:2012nd,Krajewski:2012aw}. Its support at small deficit angle leads to the so called, \emph{flatness} of spinfoam model. The flatness has been suspected to be a bad property since it seemed to imply that the semiclassical geometries from spinfoam amplitude was always flat. However the analysis here shows that the flatness property is actually a good property of spinfoam model. Regularizing the spin-sum leads to $D_\delta$ which gives a good control of the small deficit angle. The ``regularized flatness'' frees the curvature in the low curvature region and makes the simplicial geometries approximate the smooth geometries. In the high curvature region, the flatness property guarantees the strong quantum effect near curvature singularity, such that the physics is deviated away from classical gravity.

The large-$j$ spinfoam and small-$j$ spinfoam of low and high curvature regions may be viewed as two \emph{phases} of spinfoam model. The continuation of spinfoam from low to high curvature regions may be understood as the phase transition from large-$j$ phase to small-$j$ phase. The spinfoam model behaves differently in two different phases. In large-$j$ phase, the vacua of spinfoam are the semiclassical 4d simplicial geometries, on which the spinfoam degrees of freedom are the excitations producing $1/j$-corrections. In small-$j$ phase, the vacuum of spinfoam is the state with vanishing spin everywhere (no-geometry state or the so called Ashtekar-Lewandowski vacuum). The spinfoam degrees of freedom on this vacuum are the spin and intertwiner excitations. The phases proposed here might have the relation with the recent works \cite{Bahr:2016hwc,Gielen:2016uft}.

It is useful to find an order parameter in order to understand the phase transition between large-$j$ and small-$j$ phases. In Section \ref{phases}, we proposes a candidate of order parameter, being the imaginary part $\mathrm{Im}\langle j\rangle$ of the expectation value of the spin $j$. The discussion in Section \ref{phases} suggests that $\mathrm{Im}\langle j\rangle\ll 1$ in the large-$j$ phase while it should be finite in the small-$j$ phase. In Section \ref{correlation}, we analyze the correlation function of two spins located at different triangles. We find that in the large-$j$ phase, the pair of spins has a strong and long-range correlation. The correlation function doesn't decay even for a pair of spins located far away.

In this paper, the understanding of the phase and their transition is qualitative. Given a space-time with curvature singularity, it is not clear at the moment where precisely the phase transition occurs in the space-time. However the analysis suggests that the transition between large-$j$ and small-$j$ phases may happen at certain place where the curvature is still much lower than the Planckian curvature, i.e. at $L\gg \ell_P$. So the small-$j$ phase may not only cover the Planckian curvature region, but also cover a much larger domain. This effect may be resulting from the large number of spinfoam degrees of freedom on the refined triangulation, which accumulates and produces a strong quantum effect. A more quantitative understanding of the phase transition is a research undergoing currently, whose result will be reported elsewhere.

\section{Lorentzian Spinfoam Amplitude and Large Spin Asymptotics}\label{sec1}

Our analysis here is based on the Lorentzian spinfoam amplitude proposed by Engle-Pereira-Rovelli-Livine (EPRL) \cite{Engle:2007wy}. The spinfoam amplitude defined on a simplicial complex $\mathcal{K}$ can be written in an integral representation \cite{Han:2013gna}
\begin{eqnarray}
Z(\mathcal{K})&=& \sum_{j_f} \prod_f\dim(j_f)\,A_{j_f}(\mathcal{K})\nonumber\\
&=& \sum_{j_f} \prod_f\dim(j_f)\,\int_{\SLtc} \prod_{(v,e)} \dd g_{ve} \int_{\mathbb{CP}^1}\prod_{v\in\partial f} \dd z_{vf}~ \ex^{S[j_f, g_{ve}, z_{vf}]}\label{spinfoam}
\end{eqnarray}
The labels $v$, $e$ and $f$ are 4-simplices, tetrahedra and triangles in the complex $\mathcal{K}$, or vertices, dual edges and dual faces in the dual 2-complex $\mathcal{K}^*$, respectively.  Spin $j_f$ labels $\SU(2)$ irreps associated to each triangle $f$. $g_{ve}$ is an $\SLtc$ element associated to each half-edge $(v,e)$. $z_{vf}$ is a 2-component spinor (modulo complex scaling) associated to each vertex $v$ at the boundary of the dual face $f$. The spinfoam action $S[j_f, g_{ve}, z_{vf}]$ is written as
\begin{equation}
S[j_f, g_{ve}, z_{vf}]\equiv \sum_{(ef)} j_f \left(\ln \frac{\mean{Z_{vef}, Z_{v'ef}}^2}{\mean{Z_{vef}, Z_{vef}}\mean{Z_{v'ef}, Z_{v'ef}}}+\ii\g \ln\frac{\mean{Z_{vef}, Z_{vef}}}{\mean{Z_{v'ef}, Z_{v'ef}}}\right)
\end{equation}
where $Z_{vef}\equiv g_{ve}^{\dag}z_{vf}$, $\langle,\rangle$ is an $\SU(2)$ invariant Hermitian inner product between spinors, and $\gamma\in\R$ is the Barbero-Immirzi parameter. 

The asymptotic behavior of the partial amplitude $A_{j_f}(\mathcal{K})$ has been studied in the large-$j$ regime \cite{Conrady:SL2008,Barrett:2009mw,Han:2011rf,Han:2011re,Han:2013gna}. The spins $j_f \equiv  J k_f$ scales uniformly large for all triangles $f$ as $ J\gg 1$. Here $J$ is introduced as the mean value of spins on $\mathcal{K}$. The stationary phase analysis can be employed to study the asymptotic behavior of $A_{j_f}(\mathcal{K})$ since $S$ is linear to $j_f$. The leading contribution of $A_{j_f}(\mathcal{K})$ in large-$j$ comes from the critical configurations, i.e. the solutions of $\re S=0$ and $\delta_g S=\delta_z S=0$. It turns out that generically once a critical configuration is given, a Lorentzian simpicial geometry can be reconstructed on $\mathcal{K}$  (we assume the geometry is non-degenerate), described by the edge lengths together with some signs labelling the orientations. Here the orientations include both the 4d spacetime orientation and time orientation \cite{Han:2011re,Han:2013gna}.

In the following discussion, we consider the Lorentzian geometries reconstructed from the spinfoam critical configurations, which are globally oriented and time-oriented. The leading contribution to $A_{j_f}(\mathcal{K})$, coming from a spinfoam critical configuration, gives the Regge action (discrete Einstein-Hilbert action) of 4d gravity, i.e. 
\be
A_{j_f}(\mathcal{K})\sim \exp \bigg(\ii J\sum_f\g k_f \Theta_f+\cdots \bigg)=\exp \left(\frac{\ii}{\ell_P^2} S_{Regge}+\cdots \right)
\ee
by the relation between triangle area and spin $\mathbf{a}_f\sim \g j_f\ell_P^2$. $\Theta_f$ is the deficit angle of the simplicial geometry determined by the critical configuration, which encodes the curvature of the reconstructed spacetime.

``$\cdots$'' in the above asymptotic formula stands for the $\ln J$ and $1/J$ corrections. $\ln J$ correction relates to the determinant of Hessian matrix $H_{ij}(x)=\partial_i\partial_jS(x)$ ($x^i$ denotes the spinfoam variables $g_{ve},z_{vf}$). For an integral of type $\int\dd^n x\, u(x)\,e^{J S(x)}$ ($u(x)$ is a smooth function, and corresponds to the integration measure in $Z(\mathcal{K})$), the correction of order $1/J^s$ is given by 
\be
i^{-s}\sum_{l-m=s}\sum_{2l\geq 3m}\frac{2^{-l}}{l!m!}\left[\sum_{i,j=1}^nH^{-1}_{ij}(x_0)\frac{\partial^2}{\partial x_i\partial x_j}\right]^l\left(g_{x_0}^m u\right)(x_0)\label{Lsu}
\ee
where the function $g_{x_0}(x)$ is given by $g_{x_0}(x)=S(x)-S(x_0)-\frac{1}{2}H_{ij}(x_0)(x-x_0)_i(x-x_0)_j$. When the triangulation is refined, the number of spinfoam variables $g_{ve},z_{vf}$ increases. Then there will be a large number of terms contributing the above sum $\sum_{i,j=1}^n$. It is likely that the above $1/J^s$ correction becomes large when the triangulation is refined. So $J$ should also increase while the triangulation is refined, in order to suppress the $1/J^s$ correction and keep the Regge action as the leading term.

\section{Spin-Sum, $\ii\varepsilon$-Regularization, and Small Deficit Angle}\label{sec1.5}

The semiclassical analysis of the full spinfoam amplitude $Z(\mathcal{K})$ is more subtle once the sum of $j$ is taken into account. A naive semiclassical analysis leads to the so called the ``flatness'' of the spinfoam amplitude. Let us consider the sum of spins only in the large spin regime. We may approximate the spin-sum in $Z(\mathcal{K})$ as an integral 
\begin{equation}\label{eq:naiveInt}
Z(\mathcal{K})\sim 4^{N_f}J^{2N_f}\int\prod_f k_f \dd k_f ~\int_{\SLtc} \prod_{(v,e)} \dd g_{ve} \int_{\mathbb{CP}^1}\prod_{v\in\partial f} \dd z_{vf}~ \ex^{J\sum_f k_f F_f[g_{ve},z_{vf}]}
\end{equation}
where the spinfoam action $S$ is rewritten as $J\sum_f k_f F_f[g_{ve},z_{vf}]$. When $J\gg 1$, if the stationary phase approximation was employed, the amplitude would be controlled by the data $(j^0_f,g^0_{ve},z_{vf}^0)$ which were the solutions of $\re S = 0$ and $\delta_k S = \d_g S = \d_z S = 0$. The solutions turn out to give the simplicial geometries with $\gamma \Theta_f=0$ \footnote{One might replace the spin-sum by integral using Poisson resummation formula, which led to $\gamma \Theta_f\in 4\pi \ii\Z$ \cite{Han:2013hna}.}, which seem to all correspond to the flat geometry. It seems to imply that semiclassically the amplitude would be dominated by flat geometry in 4d. This property is usually refered to as the \emph{flatness} of spinfoam amplitude \cite{Han:2013hna,Hellmann:2013gva,Bonzom:2009hw,Perini:2012nd}. 

However at the solutions corresponding to flat geometry, the Hessian matrices are degenerate, which means that the stationary phase approximation based on Gaussian type integral becomes obscure for treating the spin-sum in Eq.(\ref{eq:naiveInt}). The solutions are degenerate critical points because a flat geometry admits too many triangulations with flat 4-simplices. A pair of triangulations can be arbitrarily close to each other (according to a certain norm on the parameter space), e.g. a vertex in the triangulation can move continuously while the simplicial geometries are always flat. Each triangulation of flat geometry is a critical point for Eq.\Ref{eq:naiveInt}. When there are two arbitrarily closed critical points, the critical points are in general degenerate.  

In order to overcome the incapability of the stationary phase analysis, we have to explicitly perform the spin-sum in the spinfoam amplitude. Now we focus on a neighborhood of a large-$j$ critical configuration $(j_f^0,g^0_{ve},z_{vf}^0)$ which corresponds to a globally oriented and time-oriented Lorentzian geometry. We not only consider the integration of $g_{ve},z_{vf}$, but also take into account of the sum over the spins in the neighborhood at $j^0_f\gg1$. Schematically, we compute 
\begin{equation}\label{Zjgz}
Z_{(j^0_f,g^0_{ve},z_{vf}^0)}(\mathcal{K})= \int_{N(g^0_{ve},z_{vf}^0)}\prod_{(v,e)} \dd g_{ve} \prod_{v\in\partial f} \dd z_{vf}\,\ex^{\sum_f j^0_f F_f}\prod_f\sum_{s_f}\left(2j^0_f+1+2s_f\right)\ex^{s_f F_f}
\end{equation}
where $N(g^0_{ve},z_{vf}^0)$ is the neighborhood at $(g^0_{ve},z_{vf}^0)$. $s_f=j_f-j_f^0$ is the fluctuations of spins at the large spins $j_f^0$. 

It is interesting to understand the sum $\sum_{s_f}$ of the perturbations. It has to be essentially a finite sum by the following reason: The magnitude of $\{j_f\}$ introduces an energy scale to the system. Because of the LQG area spectrum, $\g j_f\ell_P^2$ is the area of each plaquette in the simplicial lattice. When we study the physics at a given energy scale, the energy scale relates to the size of the lattice plaquette, and relates to a certain magnitude of $j_f$. We only consider the fluctuation of $j_f$ which doesn't go much beyond the given scale $j^0_f$. In particular, we don't consider the deviation of $j_f$ which goes much below ${j}^0_f$ and touches the small-$j$ regime. The small-$j$ makes the LQG area closes to the Planck scale, thus is a deep quantum regime. 

The situation of spinfoam model is very different from the usual context of renormalization group in QFT. In QFT, one often integrates out the high energy modes to understand the low energy physics. But here the sum of $j_f$ is not a sum over high/low energy modes, but rather a sum over energy scales themselves. The appearance of summing over scales in the theory essentially because the theory sums all the geometries in a background independent manner. Therefore we wouldn't expect the physical theory defined at a given energy scale came from a sum over all other energy scales (because here it is not a sum over modes at scales but a sum of scales themselves). We also wouldn't expect the physical theory at a certain scale dominating the contribution in nature. So in our opinion, it doesn't make sense to ask whether the contribution from large-$j$ or any scale of $j$ should dominate the spinfoam amplitude. Here when we analyze the physics at a given energy scale (corresponding to $j^0_f$), we focus on a regime of spin-sum within a neighborhood at this scale, and ignore the contribution in $Z(\mathcal{K})$ from other scales.  

However it is not completely clear how much should be the size of the neighborhood at $j^0_f$. It is difficult to make a precise cut-off of the sum over $j_f$, to decide whether the scales are much beyond $j^0_f$ or not. Therefore instead of making a cut-off, we introduce two decaying regulators $e^{-\delta^{(1,2)}_f s_f}$ in the sum to suppress the large fluctuations, and we define a regulated distribution:
\begin{equation}
D_{\delta}(F_f)\equiv\sum_{s_f=0}^{\infty}\left(2j^0_f+1+2s_f\right)\ex^{s_f (F_f-\delta^{(1)}_f)}+\sum_{s_f=-\infty}^{-1/2}\left(2j^0_f+1+2s_f\right)\ex^{s_f (F_f+\delta^{(2)}_f)}
\end{equation}
where $\delta^{(1,2)}_f>0$. Suppose the real part of $\re F_f\in[-\delta_f,0]$ in the neighborhood $N(g^0_{ve},z_{vf}^0)$, then $\delta^{(2)}_f>\delta_f$ such that $\exp [s_f (F_f+\delta^{(2)}_f)]$ is suppressed while $s_f$ goes to $-\infty$ \footnote{Introducing two different regulators $\delta^{(2)}_f\neq \delta^{(1)}_f$ because $F_f$ is complex valued. This technical imperfection will be alleviated in the formulation of spinfoam using Chern-Simons theory \cite{SFST}.}.

Recall that we focus on the neighborhood $N(g^0_{ve},z_{vf}^0)$ in Eq.\Ref{Zjgz} because we are in the regime of large $j_f$. We can estimate the relation between $\delta_f$ or $\delta^{(2)}_f$ and the scale of $j^0_f$. Let's consider a compact neighborhood $K$ in $(g_{ve},z_{vf})$-space, which is away from the submanifold defined by $\re F_f=0$. Recall that the real part of $F_f$ is non-positive $\re F_f\leq 0$ for all $f$. Then there exists a $\delta_f>0$ such that in $K$, $\re F_f\leq -\delta_f$ at least for one $f$. It is clear that $K$ doesn't contain any critical point. Given an oscillatory integral $\int_K e^{J S}\dd\mu$ with $\re S\leq 0$ on $K$, if there is no critical point of $S$ in the integration domain $K$ \cite{Hormander}, 
\be
\left|\int_K e^{J S(x)}\dd \mu(x)\right|\leq C\left(\frac{1}{J}\right)^k\sup_K\frac{1}{\left(|S'|^2-\re S\right)^{k}}
\ee
the integral decays faster than $(1/\l)^k$ for all $k\in\Z_+$, provided that $\sup([|S'|^2+\re S]^{-k})$ is finite (i.e. doesn't cancel the $(1/\l)^k$ behavior in front). Here because in $K$, $\re F_f\leq -\delta_f$ at least for one $f$,
\be
|S'|^2-\re S\geq|S'|^2+{k}_f\delta_f \geq{k}_f\delta_f\ \ \Rightarrow\ \ \frac{1}{J^k}\sup_K\frac{1}{\left(|S'|^2-\re S\right)^{k}}\leq\frac{1}{({j}_f\delta_f)^k}.
\ee
where ${k}_f={j}_f/J$. So the integration on $K$ suppresses when 
\begin{equation}\label{eq:jdelta}
j_f\delta_f>1.
\end{equation}
It means that we can ignore the contribution from $K$ in $(g_{ve},z_{vf})$-space in Eq.\Ref{Zjgz}. Therefore as $j_f^0\gg1$, we restrict our attention to $N(g^0_{ve},z_{vf}^0)$ with $\re F_f\in[-\delta_f,0]$, when Eq.\Ref{eq:jdelta} is satisfied\footnote{The contribution from $j_f$ far away from $j_f^0$ is suppressed by the decaying regulators. So we essentially focus on a neighborhood of large $j$'s.}. In the following we set $\delta_f^{(1,2)}\sim\delta_f\sim 1/J$, while $\delta^{(2)}_f>\delta_f>1/J$.

Perform the sum of $s_f$, $D_{\delta}(F_f)$ becomes
\begin{equation}
D_{\delta}(F_f)=\frac{2j_f^0+1-2j_f^0\ex^{(F_f-\delta^{(1)}_f)/2}}{\left[1-\ex^{(F_f-\delta^{(1)}_f)/2}\right]^2}-\frac{2j_f^0+1-2j_f^0\ex^{(F_f+\delta^{(2)}_f)/2}}{\left[1-\ex^{(F_f+\delta^{(2)}_f)/2}\right]^2}
\end{equation} 
It is obvious that $D_{\delta}(F_f)$ has two series of 2nd order poles which are purely imaginary
\begin{equation}\label{eq:poles}
F_f-\delta^{(1)}_f=4\pi\ii\mathbb{Z}, \quad F_f+\delta^{(2)}_f=4\pi\ii\mathbb{Z}
\end{equation}
Since $\re F_f\in[-\delta_f,0]$ in the neighborhood $N(g^0_{ve},z_{vf}^0)$, and $\delta^{(2)}_f>\delta_f$, we have $\re F_f-\delta^{(1)}_f\leq-\delta^{(1)}_f<0$ and $\re F_f+\delta^{(2)}_f\geq \delta^{(2)}_f-\delta_f >0$. The real parts of $F_f-\delta^{(1)}_f$ and $F_f+\delta^{(2)}_f$ are not zero, which means that the poles of $D_{\delta}(F_f)$ are all falling outside of $N(g^0_{ve},z_{vf}^0)$. $D_{\delta}(F_f)$ is a smooth function in the domain of $N(g^0_{ve},z_{vf}^0)$. The implementation of $\delta^{(1,2)}_f$ might be viewed as an analog of the $\ii\varepsilon$-regularization of Feynman propagator in QFT. 

The regularized contribution $Z^{(\delta)}_{(j^0_f,g^0_{ve},z_{vf}^0)}(\mathcal{K})$ is defined from Eq.\Ref{Zjgz} by regularizing the sum over $s_f$
\be
Z^{(\delta)}_{(j^0_f,g^0_{ve},z_{vf}^0)}(\mathcal{K})=\int_{N(g^0_{ve},z_{vf}^0)}\dd g_{ve}\dd z_{vf}\, \ex^{\sum_{f} j^0_f F_f[g_{ve},z_{vf}]}\prod_{f}D_\delta\left(F_f[g_{ve},z_{vf}]\right).
\ee
Since $D_\delta(F_f)$ is a smooth function on $N(g^0_{ve},z_{vf}^0)$ and $j^0_f\gg1$, the above integral can be analyzed by the standard stationary phase approximation. There is a singe critical point $(j^0_f,g^0_{ve},z_{vf}^0)$ inside $N(g^0_{ve},z_{vf}^0)$. We have the asymptotic formula with the Regge action as the leading effective action
\begin{equation}
Z^{(\delta)}_{(j^0_f,g^0_{ve},z_{vf}^0)}(\mathcal{K})\sim \ex^{\ii\sum_f\g j_f^0\Theta^0_f}\prod_f D_{\delta}\left(\ii\g\Theta_f^0\right)\left[1+O\left(J^{-1}\right)\right]\label{ZeD}
\end{equation}
where $\Theta_f^0$ is the deficit angle reconstructed from the critical configuration.

At the critical point $(j_f^0,g^0_{ve},z_{vf}^0)$, $F_f$ takes purely imaginary value $F_f=\ii\g\Theta^0_f$ where $\Theta^0_f$ is the deficit angle at $f$. From the expression of $D_\delta(F_f)$, it is clear that $D_\delta(i\g\Theta^0_f)$ becomes large when $\ii\g\Theta^0_f$ approach close to one of the poles $F_f=4\pi \ii\Z \pm \delta^{(1,2)}_f$, although the poles have been regularized away from the purely imaginary axis. Here we are not interested in the poles $F_f=4\pi \ii n_f \pm \delta^{(1,2)}_f$ with $n_f\neq 0$, because the critical points $(j_f^0,g^0_{ve},z_{vf}^0)$ close to these poles doesn't correspond to a proper simplicial geometry, in the sense that $\g\Theta^0_f$ close to $4\pi n_f$ $(k_f\neq0)$ implies a conical singularity located at $f$, whose physical meaning is unclear. We expect that the appearance of $4\pi n_f$ poles $(n_f\neq0)$ is an artifact of $Z(\mathcal{K})$ being a discrete theory from starting point. For example, unphysical poles of momenta in principle also appear in lattice-field-theory propagators, which is an analog to $4\pi n_f$ poles $(n_f\neq0)$ here. But the integration of momenta in lattice field theory is only over the Brillouin zone where only the physical pole is relevant.

Now we focus on the neighborhoods at the poles $F_f=\pm \delta^{(1,2)}_f$. $D_\delta(i\g\Theta^0_f)$ in the asymptotic formula Eq.\Ref{ZeD} implies that the critical points $(j^0_f,g^0_{ve},z_{vf}^0)$ with small deficit angle $\Theta^0_f\ll1$ contribute much greater than other critical points. The deficit angle relates the lattice spacing $a$ and the mean curvature radius $L$ of the geometry by \cite{FFLR}
\be
\Theta^0_f\sim\frac{a^2}{L^2}\left[1+o\left(\frac{a^2}{L^2}\right)\right]\label{epsa}
\ee
Therefore when $L$ is fixed, the simplicial geometries close to the continuum limit contribute to $Z(\mathcal{K})$ much more than other simplicial geometries. $D_\delta$ coming from spin-sum forces $(j^0_f,g^0_{ve},z_{vf}^0)$ to satisfy 
\be
a^2\ll L^2,
\ee 
in order to have nontrivial contribution to the spinfoam amplitude. The explicit behavior of $D_\delta\left(i\g\Theta^0_f\right)$ as $\Theta^0_f\ll1$ is 
\begin{eqnarray}
D_\delta\left(i\g\Theta^0_f\right)&=&\frac{4 j^0_f\left(\delta^{(1)}_f+\delta^{(2)}_f\right)}{\left(\g\Theta^0_f+i\delta^{(1)}_f\right)\left(\g\Theta^0_f-i\delta^{(2)}_f\right)}+\frac{8i\g\Theta^0_f\left(\delta^{(1)}_f+\delta^{(2)}_f\right)+4\left[(\delta^{(1)}_f)^2+(\delta^{(2)}_f)^2\right]}{\left(\g\Theta^0_f+i\delta^{(1)}_f \right)^2\left(\g\Theta^0_f-i\delta^{(2)}_f\right)^2}\nonumber\\
&&+ \text{regular in }\Theta^0_f.\label{explicitD}
\end{eqnarray}
where we see $D_\delta$ is much greater when $\Theta^0_f\ll1$ than when $\Theta^0_f$ is finite. 

The set-up $\delta_f^{(2)}>\delta_f$ implies that the removal of regulator $\delta^{(2)}_f$ has to be done together with large-$j$ limit, by Eq.\Ref{eq:jdelta}. As $\delta^{(1,2)}_f$ become small, the nontrivial contribution of $D_\delta$ comes from small deficit angle $|\g\Theta^0_f|\leq \delta^{(1,2)}_f$, and $D_\delta$ behaves as
\be
D_\delta\sim \frac{4 j^0_f}{\delta^{(1,2)}_f}+\frac{2i\g+1/2}{(\delta^{(1,2)}_f)^2}.
\ee
The relation $j^0_f\delta_f>1$ is now equivalent to 
\be
\frac{a^4}{\ell_P^2 L^2}\sim|\g j^0_f\Theta^0_f|>1, \label{ellL}
\ee 
if we identify $a^2\sim \g j_f^0\ell_P^2$ and $a^2/L^2\sim\Theta^0_f\sim \g^{-1}\delta_f$. In this regime, when the number of $f$ is large in $\mathcal{K}$, the effective action in Eq.\Ref{ZeD} 
\be
\sum_f\g j_f^0\Theta^0_f=\frac{1}{\ell_P^2}\sum_f\mathbf{a}_f^0\Theta^0_f\gg 1,
\ee
and gives a rapid oscillating exponential, unless the Regge equation of motion is satisfied such that Regge action $\sum_f\mathbf{a}_f^0\Theta^0_f$ vanishes.  

Here we see that the effect of $D_\delta$ in the asymptotics Eq.\Ref{ZeD} is to suppress the contributions from the critical points $(j^0_f,g^0_{ve},z_{vf}^0)$ whose deficit angles $\Theta^0_f$ are not small. We know that a $(j^0_f,g^0_{ve},z_{vf}^0)$ with non-small $\Theta^0_f$ corresponds to a simplicial geometry which doesn't approximate any smooth geometry because of Eq.\Ref{epsa}. By summing over $j_f$, the appearance of $D_\delta$ in the asymptotics selects only the simplicial geometries which are good approximation to the smooth geometries and suppresses the rest. As a result, only those $(j^0_f,g^0_{ve},z_{vf}^0)$'s with $\Theta^0_f\ll 1$ being good approximation of smooth geometries essentially make the contributions to the semiclassical asymptotics of $Z(\mathcal{K})$. As $\delta^{(1,2)}_f\rightarrow 0$, $D_\delta$ pushes the critical points $(j^0_f,g^0_{ve},z_{vf}^0)$ contributing $Z(\mathcal{K})$ to approach the smooth geometries (approximate the smooth geometries arbitrarily well), when the simplicial complex is also refined accordingly at the same time. At each $Z_{(j^0_f,g^0_{ve},z_{vf}^0)}(\mathcal{K})$ 
\be
Z_{(j^0_f,g^0_{ve},z_{vf}^0)}(\mathcal{K})\sim \ex^{i\sum_{f} \g {j}^0_f \Theta^0_f}\simeq \ex^{\frac{i}{\ell_P^2}\int\dd^4 x\, \sqrt{-g^0}R^0}
\ee
where the Regge action approaches the Einstein-Hilbert action evaluated at the corresponding smooth geometry. Because of Eq.\Ref{eq:jdelta}, $\delta^{(1,2)}_f\rightarrow 0$ has to be combined with $j^0_f\rightarrow \infty$, i.e. the continuum limit and large-$j$ limit are taken at the same time. 

Note that even if the requirement Eq.\Ref{eq:jdelta} is alleviated (e.g. in the Chern-Simons formalism \cite{SFST}), one may still need to increase $j_f$ at the same time as refining the triangulation. The reason is that when the triangulation is refined, the $1/j_f$ quantum corrections may become larger, since more degrees of freedom are summed. Then $j_f$ may have to increased to suppress the quantum corrections, as mentioned at the end of Section \ref{sec1}.

It is important to emphasize that when we set the scale of the theory to be in the large-$j$ regime, the (regularized) spin-sum of the spinfoam amplitude forces the spinfoam critical configurations to correspond to simplicial geometries with small deficit angle $\Theta_f^0\ll 1$, i.e. the resulting simplicial geometries have to satisfy
\be
\ell_P^2\ll a^2\ll L^2,\label{laL}
\ee
in order to have the nontrivial contribution to the spinfoam amplitude.

\section{High Curvature Leads to Small Spins}\label{sec2}

The previous discussion focuses on the large-$j$ regime of spinfoam amplitude. Eq.\Ref{laL} means that the spinfoam configuration $(j_f^0,g_{ve}^0,z_{vf}^0)$, which contributes nontrivially to $Z(\mathcal{K})$, is a semiclassical space-time with a relatively low curvature. In this section, we consider the behavior of spinfoam for a space-time containing a high curvature region. A typical example is the space-time with curvature singularity. 

In the following we consider the spinfoam configuration $(j_f^0,g_{ve}^0,z_{vf}^0)$ have the following properties: (1) It has a subset of data $(\bar{j}_f^0,\bar{g}_{ve}^0,\bar{z}_{vf}^0)\subset (j_f^0,g_{ve}^0,z_{vf}^0)$ being large-$j$ and critical. The subset of data correspond to a low curvature region of a space-time geometry satisfying Eq.\Ref{laL} and Einstein equation; (2) In addition to the low curvature region, the space-time geometry relating to $(\bar{j}_f^0,\bar{g}_{ve}^0,\bar{z}_{vf}^0)$ (in its low curvature region) also has a high curvature region; and (3) $(j_f^0,g_{ve}^0,z_{vf}^0)$ should have nontrivial contribution to the full spinfoam amplitude $Z(\mathcal{K})$ (its contribution is not suppressed). We ask the following question: how does the rest of data $(j_f^0,g_{ve}^0,z_{vf}^0)\setminus(\bar{j}_f^0,\bar{g}_{ve}^0,\bar{z}_{vf}^0)$ behave in the high curvature region?

As an example, we may consider a Schwarzschild black hole, where the large-$j$ critical configuration $(\bar{j}_f^0,\bar{g}_{ve}^0,\bar{z}_{vf}^0)$ describes the low curvature geometry outside the event horizon. We are interested in how the data $(\bar{j}_f^0,\bar{g}_{ve}^0,\bar{z}_{vf}^0)$ continues to the high curvature region inside the event horizon, especially near the singularity.

As a quick answer, the spinfoam data $(j_f^0,g_{ve}^0,z_{vf}^0)$ describing a high curvature space-time (near classical singularity) must have small $j^0_f$, i.e. the spinfoam cannot be semiclassical in high curvature region. There are two steps of explanation:

\begin{itemize}
\item Firstly, in the high curvature region, a spinfoam data $(j_f^0,g_{ve}^0,z_{vf}^0)$ cannot satisfy both $\ell_P^2\ll a^2$ and $a^2\ll L^2$. Suppose we keep the data $(j_f^0,g_{ve}^0,z_{vf}^0)$ satisfying $a^2\ll L^2$ in high curvature region (small $L^2$), when the curvature becomes almost Planckian $L^2\sim \ell_P^2$, the lattice spacing $a$ has to approach the Planck scale $\ell_P$, violating $\ell_P^2\ll a^2$. Considering $a^2\sim\g j_f \ell_P^2$, the spin $j_f$ has to be small (we always assume $\g\sim o(1)$). Another way to conclude, if $(j_f^0,g_{ve}^0,z_{vf}^0)$ was a large-$j$ critical configuration in high curvature region, the deficit angle would violate $\Theta_f^0\ll1$. 

\item Secondly, now let's assume $(j_f^0,g_{ve}^0,z_{vf}^0)$ to be a large-$j$ critical configuration and $\Theta_f^0$ is not small. We apply the analysis in the last section to perform the sum over spins at the given scale. The regularized spin-sum results in the distribution $D_\delta(i\g \Theta_f^0)$ in Eq.\Ref{ZeD}. Therefore the large-$j$ critical configuration $(j_f^0,g_{ve}^0,z_{vf}^0)$ only gives a tiny contribution when $\Theta_f^0$ is not small. So it contradicts to our requirement that the contribution from $(j_f^0,g_{ve}^0,z_{vf}^0)$ to $Z(\mathcal{K})$ is nontrivial. Therefore the only possibility is that a spinfoam data $(j_f^0,g_{ve}^0,z_{vf}^0)$ describing the high curvature space-time region near a classical singularity is of small-$j$ and is not semiclassical. 

\end{itemize}


Therefore we conclude that approaching the high curvature region near a classical singularity, the spinfoam amplitude forces the spins $j_f$ to become small, in order that the contribution to the spinfoam amplitude is not suppressed. Thus the spinfoam amplitude are dominated by small-$j$ contributions in this region. Because of small $j_f$, the region near a classical singularity is highly quantum, and is referred to as the quantum gravity region. Here ``small-$j$'' means that $1/J$ correction in Eq.\Ref{ZeD} is not negligible, so that the semiclassical approximation breaks down. Small-$j$ may not necessarily mean that $j_f\sim O(1)$. When the triangulation is sufficiently refined, a not-so-large $j_f$ may not suppress the $1/J$ correction due to an increased number of spinfoam degrees of freedom, thus is still understood as small-$j$.   

It is clear that the large-$j$ approximation breaks down in the Planckian curvature region, where $L\sim\ell_P$, by the above argument. So the Planckian curvature region is necessary inside the small-$j$ regime of spinfoam. However, it is not clear where precisely in the spacetime, the large-$j$ turns to be small. If the small-$j$ region was precisely the Planckian curvature region, then $j_f$ would have to be of $O(1)$ in the small-$j$ region, by Eq.\Ref{laL} which should hold outside the small-$j$ region. But it is likely that the region where $j_f$ are small is much larger than the Planckian curvature region, due to the large number of spinfoam degrees of freedom. Indeed if we travel toward a classical curvature singularity, the above argument implies that $j_f$ decreases from the low curvature region to the high curvature region. But when we consider a refined triangulation, it corresponds to a large number of spinfoam degrees of freedom. A slight decreasing of $j_f$ may not anymore capable to suppress the $1/J$ correction. It is likely that we arrive the small-$j$ region far before we approach the Planckian curvature region. This phenomena may relate to the results in \cite{Haggard:2014rza}, where the author expect the quantum effect may even appear slightly outside the event horizon.


In the small-$j$ regime of spinfoam, the semiclassical relation between spinfoam configuration and discrete geometry is broken down. From spinfoam point of view, the notion of space-time geometry becomes invalid and corrected by large quantum fluctuations ($1/J$ corrections). To understand the dynamics of this regime, one should study the full nonperturbative behavior of the spinfoam amplitude $Z(\mathcal{K})$ in Eq.\Ref{spinfoam}. The nonperturbative spinfoam amplitude is well-defined and of nice properties \cite{Han:2010pz,Haggard:2014xoa,Haggard:2015yda,Han:2015gma,Rovelli:2010vv,Rovelli:2014ssa}.

We again consider a Schwarzschild black hole space-time. The space-time corresponds to a spinfoam critical configuration $(\bar{j}_f^0,\bar{g}_{ve}^0,\bar{z}_{vf}^0)$ in the low curvature region, satisfying Eq.\Ref{laL}.

\begin{figure}[htbp!]
\centering 
\includegraphics[width=\textwidth]{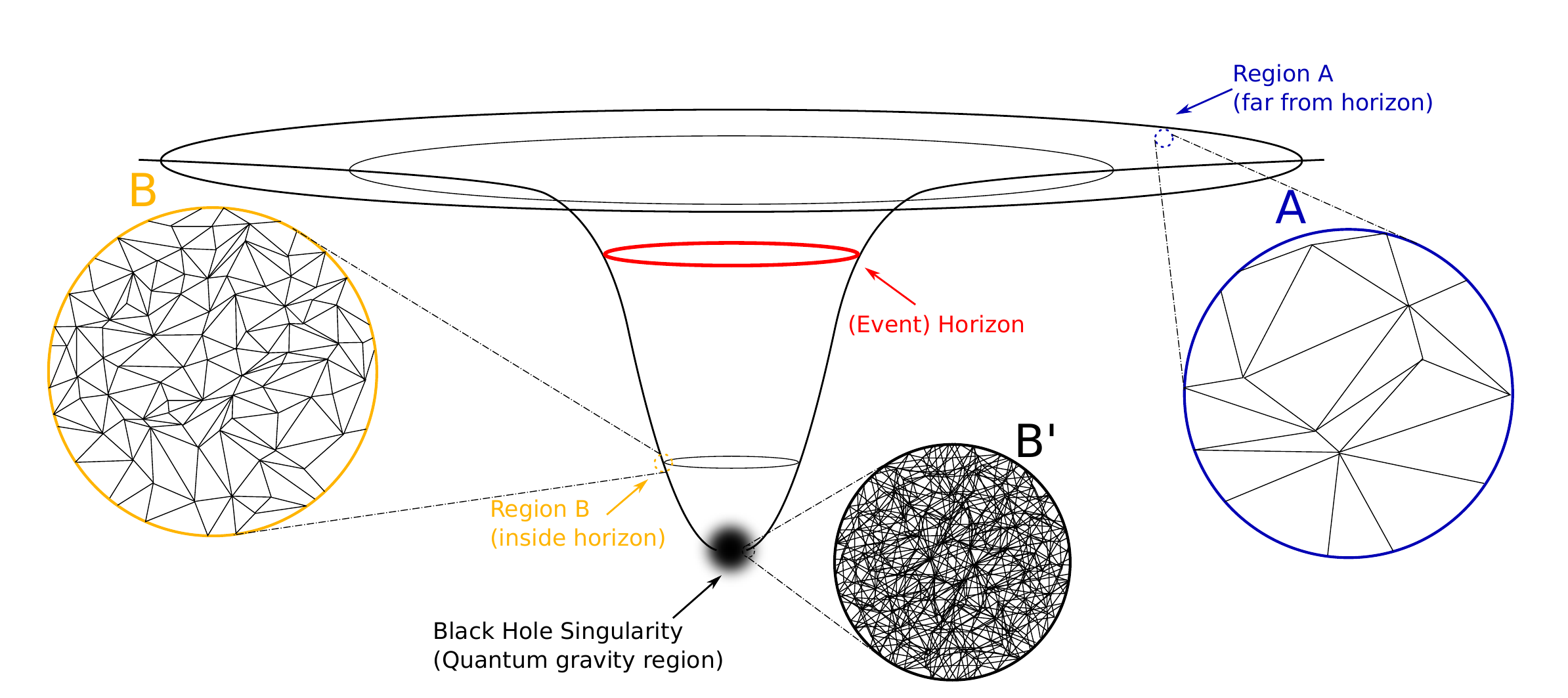}
\caption{\label{fig:i} A Schwarzschild space-time. Region A is a sub-space-time which is far from the horizon. Region B is a sub-space-time which is inside the horizon. Region B$'$ is a sub-space-time which is near the quantum region of the black hole singularity.}
\end{figure}

As shown in fig.(\ref{fig:i}), the Schwarzschild space-time is naturally divided into three regions: black hole singularity (Planckian curvature region), region inside and near the (event) horizon, region outside and far from the horizon. If we pick up a point and its neighborhood in the region far away from the horizon, the curvature in this sub-region (denoted by Region A) is small, as we learned from the Schwarzschild metric, when radius coordinate $r$ is much larger than the Schwarzschild radius $r_s = 2 G M/c^2$, the Schwarzschild metric becomes Minkowski metric with an order $(r_s/r)$ correction. So in this Region A, the mean curvature radius $L_A$ is large, and thus the average deficit angle $|\Theta^o|\ll 1$. This region corresponds to a large-$j$ spinfoam critical configuration $(\bar{j}_f^0,\bar{g}_{ve}^0,\bar{z}_{vf}^0)$ satisfying Eq.\Ref{laL}. Because of large-$j$ the triangle areas of the triangulation can be relatively large but still satisfy Eq.\Ref{laL}.

Region B denotes a sub-region inside the horizon (approaching the near-singularity region), where the background mean curvature becomes larger. The Kretschmann invariant for Schwarzschild metric at coordinate radius $r$ is $R^{\mu\nu\rho\sigma}R_{\mu\nu\rho\sigma}=12 r_s^2/r^6$, where $R_{\mu\nu\rho\sigma}$ is Riemann curvature tensor. Since Riemann curvature is scaling as the inverse of the square of the mean curvature radius $L_B$ of Region B, i.e. $R_{\mu\nu\rho\sigma}\sim L_B^{-2}$, $L_B^2$ behaves as
\begin{equation}\label{eq:Lb}
L_B\sim r \sqrt{\frac{r}{r_s}}, \quad r<r_s
\end{equation}
The background deficit angle $\Theta^0$ in Region B scales as $|\Theta^0|\sim a^2 r_s^2/r^6$.

The quantum effect (organized by $1/J$ corrections) cannot be neglected when the curvature approaches Planckian in Region B$'$, i.e. the mean curvature radius is of order of Planck length $L\sim\ell_p$. The spins $j_f$ become small in this region. However it is likely that the small-$j$ region may not only cover Region B$'$, but also cover a part of Region B or possibly even entire Region B, because of the large number of degrees of freedom on a refined triangulation. Traveling from Region A to Region B and B$'$, the quantum area $j_f$ becomes smaller. So the size of each 4-simplex in the triangulation shrinks while approaching the Planckian curvature region.

We can estimate the radius of the Planckian curvature region where the quantum gravity effect is clearly strong. By Eq.(\ref{eq:Lb}), the minimal radius is
\begin{equation}
r_p\sim \sqrt[3]{\frac{r_s}{\ell_p}} \ell_p=\sqrt[3]{\frac{2\hbar G^2 M}{c^5}}
\end{equation} 
When $M$ is the mass of the sun $M_{\odot}$, then $r_p$ is
\begin{equation}
r_p\sim \sqrt[3]{\frac{2\hbar G^2 M_{\odot}}{c^5}}\sim 10^{-22}m \gg \ell_p.
\end{equation}
We may consider an extreme case where the small $j_f$ are of $O(1)$ in the Planckian curvature region. The average area $a^2$ of the triangle is of order $\g^{-1/3}\ell_p^2$. The ratio between $r_p^2$ and $a^2$ is
\begin{equation}
\frac{r_p^2}{a^2} \sim \left(\frac{\g r_s^2}{\ell_p^2}\right)^{\frac{1}{3}}\sim \g^{\frac{1}{3}}\times 10^{25} 
\end{equation}
Taking $\g\sim o(1)$, 
we then get
\begin{equation}
\frac{r_p^2}{a^2} \sim 10^{25}, \quad  J\sim o(1),\quad \#\text{ of Simplices}=10^{50}
\end{equation}
It means that the minimal size of the quantum gravity region with the mass of the sun may be of the order $10^{50}$ bigger than the average size of the simplicies that construct Region B$'$. It means that the quantum regions is made by a large number of small quantum simplicies. So the spinfoam is highly refined in the region where quantum gravity effect is strong. 

Note that in \cite{rovelli2014planck} there has been two estimations for the radius of the quantum region (Planck star) by different argument, our estimation coincides with the rough one there. Here we still keep the Schwarzschild metric as the background space-time for high curvature region. However, in the quantum gravity region, the notion of metric is actually ill-defined. Schwarzschild metric needs to be corrected in the Region B$'$. The above discussion about Schwarzschild is a rough estimation. The numbers computed above should be corrected when it is derived in a more rigorous way, which involves a better computation using spinfoam model in small-$j$ regime.

\section{On Large Spin and Small Spin Phases, and Order Parameter}\label{phases}

As it has been shown in the above, the physics of space-time near the curvature singularity is described by the spinfoams whose spins are small. Far away from the singularity, the space-time is semi-classical and of low curvature. The corresponding spinfoam configurations are of large spins. It is clear that the spins $j_f$ are summed in spinfoam amplitude $Z(\mathcal{K})$. So the small or large spin mentioned above means the spin-sum is effectively carried out in the small or large spin regime. 

It is intuitive to consider the high curvature and low curvature regions as two different \textit{phases} of small-$j$ and large-$j$. The different phases relate to the different vacua of spinfoam model. The low curvature region is the vacuum of spinfoam being the large-$j$ critical configuration $(\bar{j}_f^0,\bar{g}_{ve}^0,\bar{z}_{vf}^0)$ satisfying Eq.\Ref{laL}. The spinfoam degrees of freedom are the excitations on $(\bar{j}_f^0,\bar{g}_{ve}^0,\bar{z}_{vf}^0)$ producing $1/J$-corrections of spinfoam amplitude. The high curvature region has the vacuum state with vanishing spin everywhere (no-geometry state or the so called Ashtekar-Lewandowski vacuum). The spinfoam degrees of freedom on this vacuum are the spin and intertwiner excitations. 

Finding an order parameter is usually helpful to understand the phases, as well as the transition between them. Here we find a candidate of order parameter to be the imaginary part of $j_f$ expectation value, $\mathrm{Im}\langle j_f\rangle$, which is expected to behave differently in difference phases. Firstly we consider the large-$j$ phase. Instead of perform the sum over spins as showing in Section 2, we integrate the group elements $g_{ve}$ and spinors $z_{vf}$ in the first place. Because the computation is in the large-$j$ regime, the integration of $g_{ve}$ and $z_{vf}$ can be performed by using the saddle point approximation in the large-$j$ limit. As shown in \cite{Han:2013hna}, the spinfoam amplitude expanding at a low curvature critical configuration $(j^0_f,g^0_{ve},z_{vf}^0)$ can be written as an effective partition function of a spin system. The amplitude is 
\begin{equation}
Z_{(j^0_f,g^0_{ve},z_{vf}^0)}(\mathcal{K})= \sum_{\{j_f\}}\left(2j_f+1\right)\exp I_{\mathcal{K}}[j_f]\label{spinsystem}
\end{equation}
where $I_{\mathcal{K}}[j_f]=I_{\mathcal{K}}[Jk_f]$ is the effective action. Define new variables $\kappa_f\equiv k_f-k_f^0$ and expand $I_{\mathcal{K}}[Jk_f]$ around $\kappa_f=0$, then the effective action is obtained as
\begin{equation}
I_{\mathcal{K}}[Jk_f]=J\left(I_0+I_1^f\kappa_f+I_2^{ff'}\kappa_f\kappa_{f'}+O\left(\kappa^3\right)\right)
\end{equation}
The first three coefficients are computed in \cite{Han:2013hna} to the leading order in $1/J$:
\begin{equation}
I_0=\ii\g k_f^0\Theta_f^0,\quad I_1^f=\ii\g \Theta^0_f-\delta_f^{1,2}, \quad I_2^{ff'}=\frac{2(1+2\ii\g-3\g^2-2\ii\g^3)}{5+2\ii\g} n^T_{ef}X_e^{-1}n_{ef'}
\end{equation}
where $\Theta_f^0$ is the deficit angle given by the critical configuration; $n_{ef}$ is the unit 3-vector normal determined by $(j^0_f,g^0_{ve},z_{vf}^0)$ which is the normal vector of the triangle $f$ in the frame of tetrahedron $e$ \cite{Livine:2007vk}. The matrix $X_e$ is $X_e^{ij}\equiv \sum_f k_f(-\delta^{ij}+n_{ef}^in_{ef}^j)$. The expectation value of $\langle j_f\rangle=j_f^0+\langle \kappa_f\rangle$ to the leading order in $1/J$ can be obtain by the equation of motion of $I_{\mathcal{K}}$ \cite{Han:2013hna}.
\be
\langle \kappa_f\rangle\sim \frac{1}{2} \sum_{f'}\left(I_2^{-1}\right)_{ff'}\big(\ii\g \Theta^0_{f'}-\delta_{f'}^{1,2}\big)+ O\left((\ii\g \Theta^0_f-\delta_f^{1,2})^2\right).
\ee
The above is an expansion in the low curvature regime, where $\g \Theta^0_{f'}\sim\delta_{f'}^{1,2} \sim 1/J$. Therefore in the low curvature regime, 
\be
\mathrm{Im}\langle j_f\rangle=\mathrm{Im}\langle \kappa_f\rangle\sim 1/J
\ee
is suppressed by large-$J$. In particular if we consider a black hole spacetime with asymptotically flat region, we can set both $\Theta^0_{f}$ and $\delta_{f}^{1,2}$ to be very small, corresponding $J$ being very large. Then $\mathrm{Im}\langle j_f\rangle$ is very small in the region. 

When we approach the high curvature regime, $j_f$ becomes small so that the $1/J$ corrections are not negligible. Then $\mathrm{Im}\langle j_f\rangle$ cannot be suppressed by $1/J$, and likely becomes a finite number. In the small-$j$ phase, we insert $j_f$ of a triangle $f$ into the integration formula Eq.\Ref{spinfoam} of $Z(\mathcal{K})$. The sum of $j_f$ is carried out in the small-$j$ regime. The expectation value of $j_f$ is written as
\be
\langle j_f\rangle =\frac{1}{Z(\mathcal{K})}\sum_{\{j_{f'}\}} \prod_{f'}\dim(j_{f'})\,j_f\int_{\SLtc} \prod_{(v,e)} \dd g_{ve} \int_{\mathbb{CP}^1}\prod_{v\in\partial f'} \dd z_{vf'}~ \ex^{S[j_{f'}, g_{ve}, z_{vf'}]}.
\ee
The integrand is a complex function, and the large-$j$ approximation breaks down in this phase. The large $1/J$ corrections suggests that $\mathrm{Im}\langle j_f\rangle$ should be generically nonzero and finite, although a mathematically rigorous proof of $\mathrm{Im}\langle j_f\rangle$ being finite (i.e. a lower bound of $\mathrm{Im}\langle j_f\rangle$) is still lacking for the small-$j$ phase. 

Therefore the above argument suggests that the quantity $\mathrm{Im}\langle j_f\rangle$ should have two different behavior in the large-$j$ and small-$j$ phases:
\begin{eqnarray}
&\mathrm{Im}\langle j_f\rangle\sim 1/J\ll1 &\quad \text{in large-$j$ phase (low curvature)}\nonumber\\
&\mathrm{Im}\langle j_f\rangle= \text{finite} &\quad \text{in small-$j$ phase (high curvature)}.
\end{eqnarray}
$\mathrm{Im}\langle j_f\rangle\ll 1$ is consistent with the large-$j$ interpretation of $j_f$ as semiclassical triangle area, while the finite $\mathrm{Im}\langle j_f\rangle$ in small-$j$ phase means that the semiclassical approximation breaks down. 

It should be noted that the above argument toward the order parameter is still at the qualitative level. The more detailed investigation is postponed in the future research.

\section{Correlation of Spins in the Large Spin Phase}\label{correlation}

The behavior of correlation functions is usually useful to understand the phases and their transition. Here we view the spinfoam amplitude as a ``statistical system'', and we study the correlation function of a pair of spins $j_f,j_{f'}$ at different locations. We find that in the large-$j$ phase (low curvature region), the correlation between spins is of long-range, i.e. no matter how ``far'' away the two different spins are separated, their correlation hardly decays.

Recall Eq.\Ref{spinsystem}, the spinfoam amplitude can be written perturbatively in the large spin regime with effective action $I_{\mathcal{K}}[Jk_f]$, where $\kappa_f$ is the perturbation of $j_f$ at $j_f^0$:
\begin{equation}
Z_{(j^0_f,g^0_{ve},z_{vf}^0)}(\mathcal{K})=\left(2J\right)^{N_f} \ex^{J I_0}\sum_{\{\kappa_f\}}\left(k_f^0+\kappa_f+\frac{1}{2J}\right)\ex^{J\left(I_1^f\kappa_f+I_2^{ff'}\kappa_f\kappa_{f'}+ O\left(\kappa^3\right)\right)}
\end{equation}
We keep the effective action to the quadratic order in the perturbation $\kappa_f$, and approximate the sum $\sum_{\{\kappa_f\}}$ by an integral. The amplitude $Z_{(j^0_f,g^0_{ve},z_{vf}^0)}(\mathcal{K})$ looks like a path integral over $\kappa_f$ with an external source $JI_1^f$. 

The (connected) correlations between two spins $j_f$ and $j_{f'}$ is computed at the leading order
\begin{eqnarray}
\langle (j_f-j_f^0) (j_{f'}-j_{f'}^0)\rangle &=& J^2\langle \kappa_f \kappa_{f'}\rangle = \frac{\partial_{I_1^f}\partial_{I_1^{f'}}Z_{(j^0_f,g^0_{ve},z_{vf}^0)}(\mathcal{K})}{Z_{(j^0_f,g^0_{ve},z_{vf}^0)}(\mathcal{K})}\nonumber \\
&\sim&  \frac{J^2}{4} I_1^{f_1}(I_2^{-1})^{f_1 f}I_1^{f_2}(I_2^{-1})^{f_2 f'}-\frac{J}{2}(I_2^{-1})^{ff'}
\end{eqnarray}
Because in the low curvature regime, $\g \Theta^0_{f'}\sim\delta_{f'}^{1,2} \sim 1/J$, i.e. $I_1^{f}\sim 1/J$, the leading contribution to the correlation function comes from the second term: 
\begin{equation}
\langle (j_f-j_f^0) (j_{f'}-j_{f'}^0)\rangle = -\frac{J}{2}(I_2^{-1})^{ff'}
\end{equation} 
The matrix elements of $I_2^{ff'}$ is non-zero only when triangles $f$ and $f'$ belong to the same tetrahedron. The non-zero elements of $I_2^{ff'}$ are mainly next to the diagonal. However the matrix is not a block-diagonal matrix. Then its inverse $(I_2^{-1})^{ff'}$ is also not block-diagonal. Moreover, the matrix elements $(I_2^{-1})^{ff'}$ are generically nonvanishing for an arbitrary pair of $f,f'$. The correlation between two spins are of long-range and strong. The magnitude of correlation function scales linearly in $J$.

Indeed, to illustrate the inverse of $I_2^{ff'}$, we consider a tridiagonal matrix (analog of $I_2^{ff'}$) and its inverse (analog of $(I_2^{-1})^{ff'}$)
\be
I=\begin{pmatrix}
a_1\ & b_1 \\
c_1\ & a_2 &\ b_2 \\
&\ c_2 & \ddots & \ddots \\
& & \ddots & \ddots &\ b_{n-1} \\
& & &\ c_{n-1} &\ a_n
\end{pmatrix}\ \ \text{and}\ \  
(I^{-1})_{ij} = \begin{cases}
(-1)^{i+j}b_i \cdots b_{j-1} \theta_{i-1} \phi_{j+1}/\theta_n &\ \text{ if } i \leq j\\
(-1)^{i+j}c_j \cdots c_{i-1} \theta_{j-1} \phi_{i+1}/\theta_n &\ \text{ if } i > j\\
\end{cases}
\ee
where $\{\theta_i\}_i$ satisfy the recurrence relation $\theta_i = a_i \theta_{i-1} - b_{i-1}c_{i-1}\theta_{i-2} \text{ for } i=2,3,\ldots,n$ with initial conditions $\theta_0 = 1$, $\theta_1 = a_1$. $\{\phi_i\}_i$ satisfy $\phi_i = a_i \phi_{i+1} - b_i c_i \phi_{i+2} \text{ for } i=n-1,\ldots,1$ with the initial conditions $\phi_{n+1} = 1$ and $\phi_n = a_n$ (see \cite{tridiagonal} for examples of symmetric tridiagonal matrix). Generically, all the matrix elements of $I^{-1}$ are nonvanishing. 

The correlation function in small-$j$ phase (high curvature region) is more difficult to compute, due to the lack of approximation scheme. However we do believe the spin-spin correlation function in small-$j$ phase should be of dramatically different behavior from it is in the large-$j$ phase, because the $1/J$ correction becomes non-negligible in small-$j$ phase. The further investigation of correlation function is postponed to the future research.

\acknowledgments

The authors acknowledge the helpful discussions with Jonathan Engle, Francesca Vidotto, Carlo Rovelli, and Zhaolong Wang, and acknowledge the comments from Daniele Oriti and Edward Wilson-Ewing. MZ acknowledges the funding received from Alexander von Humboldt Foundation. MH acknowledges the Institute of Modern Physics at Northwestern University in Xi'an, Yau Mathematical Sciences Center at Tsinghua University in Beijing, and Fudan University in Shanghai, for their hospitality during his visits. MH also acknowledges the support from the US National Science Foundation through grant PHY-1602867, and the Start-up Grant at Florida Atlantic University, USA.


\providecommand{\href}[2]{#2}\begingroup\raggedright\endgroup

\end{document}